\documentclass[aps,pre,nofootinbib,showpacs,twocolumn]{revtex4}
\usepackage{amssymb,latexsym,mathrsfs}
\usepackage{amsfonts}
\usepackage{amsmath}
\usepackage{graphicx,changes,color}

\newcommand{\beq}{\begin{equation}}
\newcommand{\eeq}{\end{equation}}
\newcommand{\beqn}{\begin{eqnarray}}
\newcommand{\eeqn}{\end{eqnarray}}
\newcommand{\bearr}{\begin{array}}
\newcommand{\enarr}{\end{array}}

\newcommand{\ra}{\rangle}
\newcommand{\la}{\langle}

\newcommand{\id}{\textrm{d}}

\def\bea{\begin{eqnarray}}
\def\eea{\end{eqnarray}}
\def\ba{\begin{array}}
\def\ea{\end{array}}
\def\n{\nonumber}

\begin{document}

\title{Phase Separation Transition in a Nonconserved Two Species Model}
\author{Urna Basu}
\affiliation{SISSA -- International School for Advanced Studies and INFN, via Bonomea 265, I-34136 Trieste}

\begin{abstract}
A one dimensional stochastic exclusion process with two species of particles, $+$ and $-$, is studied where density of each species can fluctuate but the total particle density is conserved. From the exact stationary state weights we show that, in the limiting case where density of negative particles vanishes, the system undergoes a phase separation transition where a macroscopic domain of vacancies form in front of a single surviving negative particle. We also show that the phase separated state is associated with a diverging correlation length for any density and the critical exponents characterizing the behaviour in this region are different from those at the transition line. The static and the dynamical critical exponents are  obtained from the exact solution and numerical simulations, respectively.
\end{abstract}


\maketitle

\section{Introduction}\label{sec:intro}

Driven diffusive systems have been a recent topic of interest because of their intriguing properties and varied range of applications \cite{zia}. One of the special features is that these systems can undergo phase transition, even in one spatial dimension. These nonequilibrium phase transitions are of various kinds and relevant in wide range of physical and biological systems \cite{marro}.  A prototypical model for studying these kind of driven systems in one dimension is the asymmetric exclusion process (ASEP)\cite{asep1,asep2} which describes stochastic motion of particles on a lattice with a bulk drive and hard core exclusion. It is well known that ordinary ASEP shows a phase transition on an open geometry \cite{asepexact}. However, many generalizations of ASEP with multiple species of particles, disordered hopping rates or kinetic constraints have been explored where phase transition occurs also on a ring  \cite{derrida, ABC, Kafri2003, rasep, evans, Krug, Krug2}.

Of particular interest is the exclusion process with two species of particles which has been studied with various sets of dynamical evolution rules  both with  and without density conservation. Some models in the first category show transitions to a phase separated state where the particles, irrespective of species, cluster together \cite{Kafri2003, PK}. Yet, other models have been studied where only a single `second class' particle is present; these systems often show jammed states where macroscopic number of ordinary particles are accumulated next to the second class particle \cite{derrida, Jafarpour,Jafarpour2}.
On the other hand, new phases and critical behaviour might emerge when particle conservation is broken. Examples include a
discontinuous transition to a completely vacant state from a homogeneous phase \cite{HH} and a continuous transition to a phase where the density of vacancies vanishes \cite{EvansMukamel}.

In the present work, we study a two-species exclusion process in the context of phase separation transition. The dynamics conserves the total number of particles although the number of particles for individual species can fluctuate. We show that this system undergoes a transition from a homogeneous liquid phase to a phase separated state, where the vacancies form a macroscopic domain, as the density of particles is changed. We also show that the phase separated state is always `critical' -- it is  associated with a diverging correlation length for any density beyond the critical one. The corresponding exponents turn out to be different from those on the critical line bounding this region. The sets of critical exponents characterizing the critical line and the phase separated state are obtained exactly. 

The paper is organized as follows. The model is defined in the next section with a brief discussion of its phenomenological behaviour. We use the exact solution and the matrix product form to study spatial correlations and behaviour of the system near criticality in Sec. \ref{sec:exact}. Mapping to a zero-range process is exploited to study the phase separated state from the canonical point of view.
The dynamical relaxation along with the finite size scaling behaviour is investigated in Sec. \ref{sec:dyn}. 
We conclude with some general discussions in Sec. \ref{sec:concl}.

\begin{figure}[t]
 \centering
 \includegraphics[width=8.6 cm]{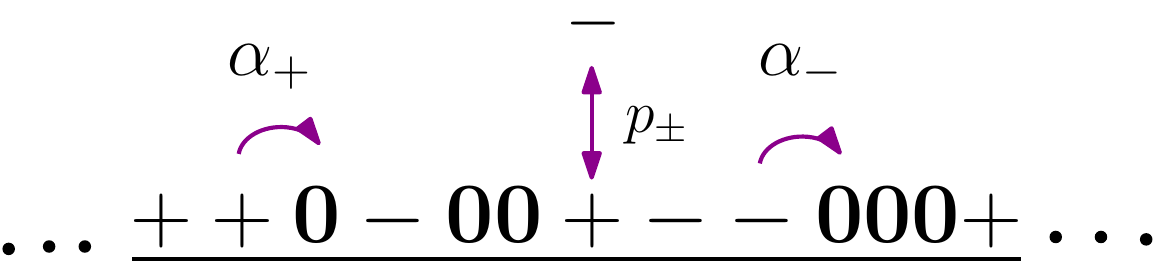}
 \caption{Schematic representation of the two species exclusion model. Positive and negative particles have different hopping rates $\alpha_\pm.$ A particle can change its type if its right neighbouring site is occupied.}
 \label{fig:cartoon}
\end{figure}

\section{The Model and Phenomenology}\label{sec:model}

The model is defined on a periodic one dimensional lattice with $L$ sites, labeled by $i=1,2, \dots L.$ Each site can be vacant or occupied by either a $+$(positive) or a $-$(negative) particle. A generic configuration of the system is thus described by the set $\{s_i\}$ where $s_i=+,-$ or $0.$ A particle attempts to hop to its right neighbouring vacant site with a rate $\alpha_{\pm}$ which depends on the type of the particle.  However, if the right neighbouring site is occupied, the particle can change its type with some rate $p_{\pm}.$ The complete dynamics can thus be summarized as,

\begin{equation}
\pm 0 \mathop{ \longrightarrow}^{\alpha_{\pm}}  0 \pm ~~~~~;~~~~~~
+ \pm \mathop{\leftrightharpoons}_{p_{+}}^{p_{-}}  -  \pm ~.
\label{eq:model}
\end{equation}
The number $M_\pm$  of particles of each type $\pm$ can fluctuate but the total number of particles $M=M_+ + M_-$ or, equivalently, the number of vacancies $N_0=L-M$ is conserved by this dynamics. We choose the conserved density of vacancies $\rho_0=N_0/L$ and the average density of negative particles $\rho_-= \la M_-\ra/L$ to be the relevant macroscopic variables describing the state of the system.  Clearly, these two quantities also fix the density of positive particles $\rho_+=1-\rho_- - \rho_0$ on the lattice.

The dynamics \eqref{eq:model} is a special case of the asymmetric exclusion process with internal degrees of freedom introduced earlier \cite{twobox}. The macroscopic densities of positive and negative particles and the nearest neighbour correlations for this system have been calculated using the exact stationary state weights. However, the possibility of a phase transition in this system has not yet been explored.

\begin{figure}[thb]
\includegraphics[width=8.8 cm]{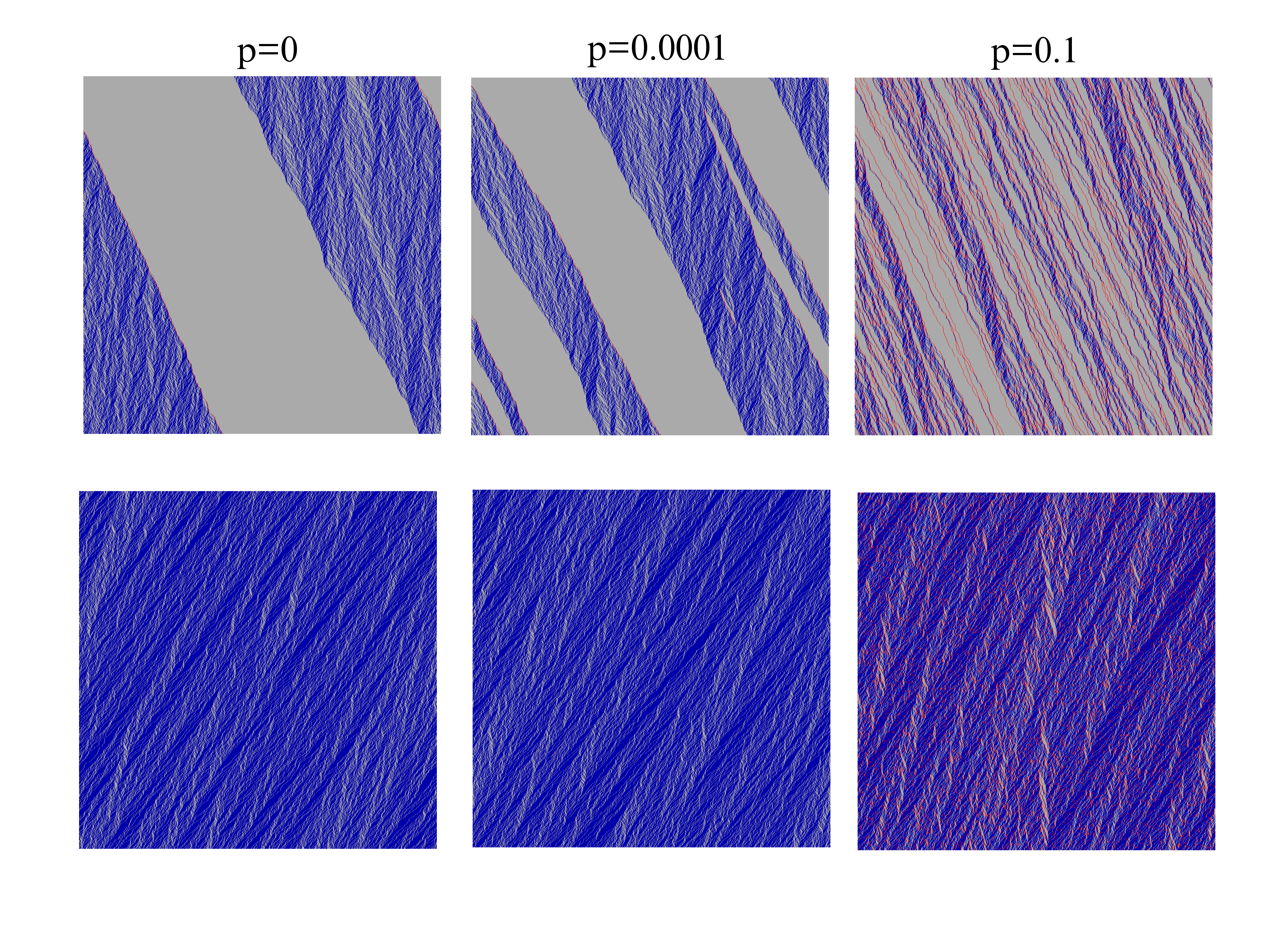}
\caption{(Color Online) Typical snapshots of time evolution of the system for $\alpha_+/\alpha_-=2$ at different points in the parameter space. The positive and negative particles are marked by red (medium gray) and blue (dark gray) pixels respectively whereas light gray pixels denote the vacancies. Upper panel corresponds to $\rho_0= 0.75$ which is larger than the critical density $\rho_c=0.5$ and the lower panel corresponds to $\rho_0=0.25 < \rho_c.$ The left, middle and right columns correspond to $p=p_+/p_-=0, 10^{-4}$ and $10^{-1}$ respectively. }
\label{fig:evolution}
\end{figure}

In this work we use the exact solution obtained in Ref.~\cite{twobox} to study a phase transition that occurs for $p_+=0.$ We will show that in this limit the system exhibits a transition from a homogeneous state to a phase separated one where a macroscopic number of vacancies cluster together in front of a negative particle. That the system admits the possibility of such a phase can be understood as follows. Let us suppose that the negative particles have slower hopping rate $\alpha_-,$ compared to positive particles, then these particles would also have longer lifetimes since the particle can change its type only when there is no vacancy in front of it.  If, additionally, the rate $p_+$ of creation of the  negative particles  is also vanishingly small, then, for a large enough density of vacancies $\rho_0,$ the system can be in a state where there are only a few negative particles with large number of vacancies clustering in front of them. From this heuristic argument one can expect that, for small $p_+,$ a transition to such a phase separated state from a liquid phase, where the particles are distributed homogeneously, can occur either by increasing the density or by decreasing the hopping rate $\alpha_-.$ Figure  \ref{fig:evolution} shows typical snapshots of time evolution of the system  in different regions of the phase space. The vertical direction represents time and it runs downwards. As expected, for small $p_+$ and high $\rho_0,$ only a microscopic number of negative particles survive and the vacancies show a tendency to conglomerate in front of these few negative particles.

A generalized version of this model, where the total number of particles is not conserved, has been studied in Ref.~\cite{HH} which also shows a phase transition. However, the corresponding state space in the non-conserved model is very different and the transition occurs from a homogeneous phase, with equal densities of both kinds of particles, to a completely empty state.

 In the following section we study the phase separation in the conserved model in detail using the exact stationary state weights.

\section{Exact Results}\label{sec:exact}

The nonequilibrium steady state of the model \eqref{eq:model} can  most conveniently be expressed in a matrix product form \cite{twobox, ozrp} which we discuss here briefly for the sake of completeness.  
Following the matrix product ansatz \cite{MPA}, the stationary state weight of a configuration $\{ s_i\}$ is written as
\bea
P(\{ s_i\}) = \text{Tr} \left[\prod_{i=1}^L X_{s_i} \right], 
\eea
where $X_{s_i}$ is the matrix corresponding to the state variable $s_i$ at the $i^{th}$ site. For this model, the ansatz demands that the matrices must satisfy the following set of algebraic relations \cite{twobox},
\bea
 \alpha_+ DA &=& -\tilde A D + A \tilde D = \tilde D A - D \tilde A \cr
 \alpha_- EA &=& -\tilde A E + A \tilde E =\tilde E A - E \tilde A \cr
 p_+ DD - p_- ED &=& -\tilde E D + E \tilde D =\tilde D D - D \tilde D \cr
 p_+ DE - p_- EE &=& -\tilde E E + E \tilde E= \tilde D E - D\tilde E, 
\eea%
where $X_+=D, X_-=E$ and $X_0=A,$ and   $\tilde D, \tilde E$ and $\tilde A$ are auxiliary matrices required to satisfy the matrix product ansatz. It turns out there is a two dimensional representation for these matrices, 
\bea
D= \begin{pmatrix} 1 & 0 \\ 1 & 0 \end{pmatrix}; 
 E= \begin{pmatrix} 0 & p \\ 0 & p \end{pmatrix}; A= \begin{pmatrix} 1 & 0 \\ 0 & \alpha \end{pmatrix}, \label{eq:dea}
\eea
 and,
\bea
\tilde D = D; \;\; \tilde E = E; \;\; \tilde A =\begin{pmatrix} 1-\alpha_+ & 0 \\ 0 & \alpha(1-\alpha_-) \end{pmatrix},  
\eea
 where $p= p_+/p_-$ and  $\alpha=\alpha_+/\alpha_-$ are the ratios of the flip rates and the hopping rates, respectively. The state of the system is completely specified by the three parameters, $\rho_0,$ $\alpha$ and $p,$  as the auxiliary matrices do not affect the stationary state weight of the configurations.

 To calculate spatial correlation functions it is convenient to work in the grand canonical ensemble where a fugacity $z,$ associated with the $A$s, fixes the average density of vacancies (i.e., $0$s). The corresponding  grand canonical partition function, for a system of size $L,$ is
\beq
Z_L= \text{Tr} [(D + E  + z A)^L] = \text{Tr}[T^L], \n
\eeq
where we have defined the transfer matrix $T= D + E  + z A= \begin{pmatrix} 1+z & p \\ 1 & p+z \alpha \end{pmatrix}.$ Thus,
$Z_L = \lambda_+^L +\lambda_-^L  \qquad \qquad $
where $\lambda_+$ and $\lambda_-$ denote the larger and smaller eigenvalue of $T,$ respectively,
\bea
\lambda_\pm &=& \frac 12 \big[1 + p + z + \alpha z \cr
&& \pm \sqrt{(1+ p + z + \alpha z)^2 - 4 z(\alpha  + p + \alpha z)} \big]. \n
\eea

The matrix product formalism allows for simple calculation of expectation values of observables. For example, the density of negative particles, in terms of the fugacity $z,$ is given by \cite{twobox}
\bea
\rho_- =\frac1L {\la M_-\ra}= \frac 1{Z_L} \text{Tr} [ET^{L-1}] \simeq \frac{p(\lambda_+ -z)}{\lambda_+(\lambda_+-\lambda_-)} \label{eq:rhom}
\eea
where the last expression is valid in the thermodynamic limit $L \to \infty.$

A typical observable used to detect a phase separation transition is the domain size of particles or, in this case, of vacancies. A domain of vacancies is defined as an uninterrupted sequence of $0$s bound by particles on both sides. In particular, we will be interested in domains which precede negative particles. The average size of such a domain per negative particle can be defined as $\la N_-\ra/ \la M_-\ra$ where  $\la N_-\ra$ is the expected value of the total number of vacancies in front of negative particles. To calculate it we note that the matrices $A$ and $E,$ as  given in Eq. \eqref{eq:dea}, have the property $EA^n = \alpha^n\left({0~~ 0 \atop 0 ~~1} \right)$ --- each sequence of $n$ uninterrupted $0$s in front of a negative particle contributes a factor $\alpha^n$ in the stationary state weight of the corresponding configuration. Hence, 
\bea
\la N_- \ra = \frac \alpha{Z_L} \frac{\id}{\id \alpha} Z_L \simeq \frac{\alpha L}{\lambda_+} \frac{\id \lambda_+}{\id \alpha}. \label{eq:Nmav}
\eea

The correlation between positive particles separated by $r$ lattice sites is given by
\bea
V_+(r) &\equiv& \la +_i+_{i+r}\ra - \la + \ra^2 \cr
&=& \frac{1}{Z_L}Tr[DT^{r-1}DT^{L-r-1}]-\rho_+^2 \cr
&=&\frac{p}{\lambda_+\lambda_-}\frac{z(\alpha-1)}{(\lambda_+-\lambda_-)^2}\left(\frac{\lambda_-}{\lambda_+}\right)^r. \label{eq:V+}
\eea
Again, the last expression is obtained in the thermodynamic limit. The correlation decays exponentially with the distance $r$ and the correlation length 
\bea
\xi=-{\left(\log \left|\frac{\lambda_-}{\lambda_+}\right| \right)}^{-1}. \label{eq:xi}
\eea
The above form for the correlation length is typical of systems with a two dimensional matrix representation
and appears in all other correlation functions. For example, the correlation between positive and negative particles separated by a distance $r,$
\bea
V_{+-}(r) &\equiv & \la +_i-_{i+r}\ra - \la + \ra \la - \ra \cr
&=& -\frac{p(\lambda_+ - z)}{\lambda_+\lambda_-}\frac{(\lambda_+ - \alpha z)}{(\lambda_+-\lambda_-)^2}  \left(\frac{\lambda_-}{\lambda_+}\right)^r. \label{eq:Vpm}
\eea
 
\begin{figure}[th]
 \centering
 \includegraphics[width=5.8 cm]{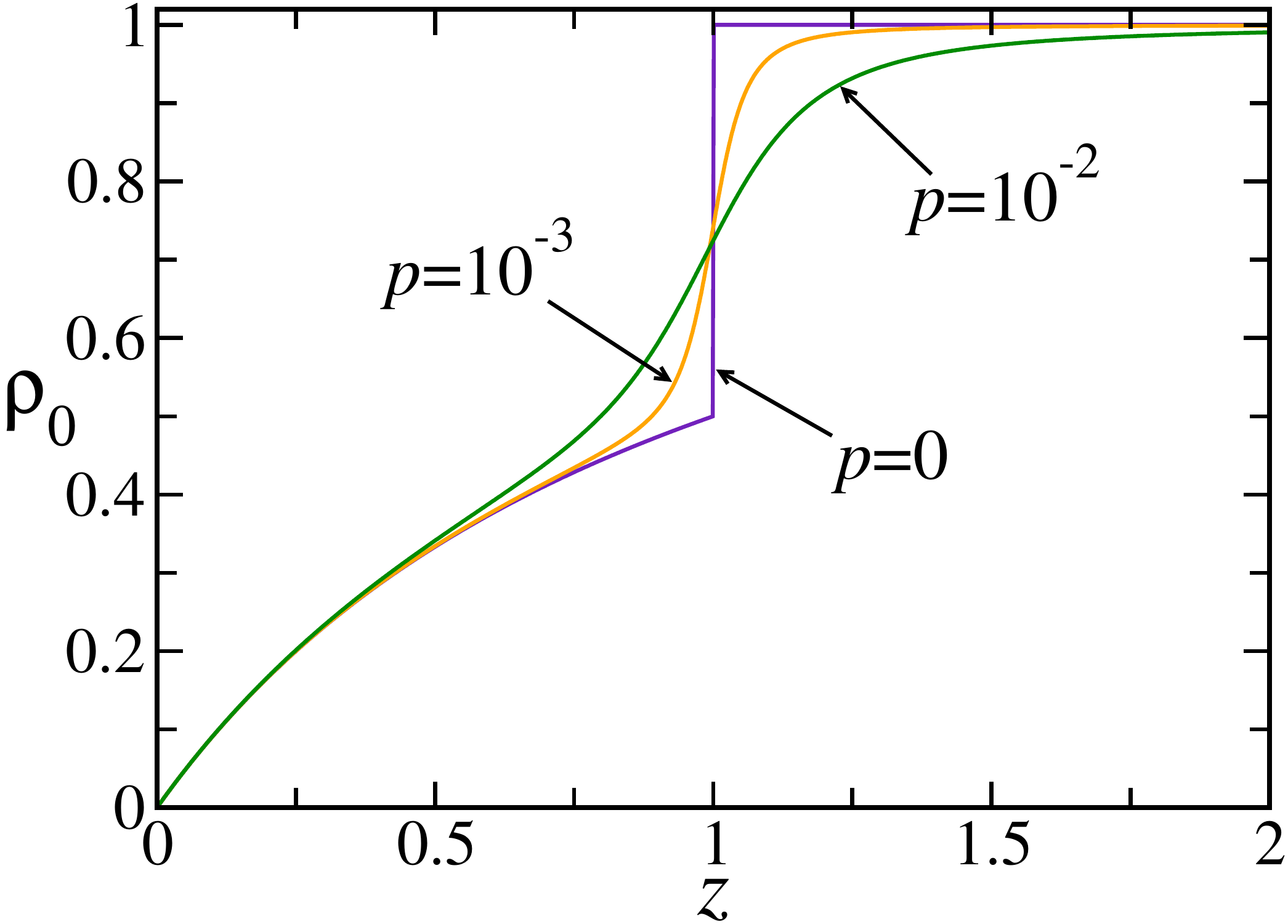}
 \caption{(Color Online) Density $\rho_0(z)$ versus the fugacity $z$ for different values of $p=0,10^{-3}$ and $10^{-2}$ for $\alpha=2.$ The density shows a discontinuous jump at $z^*=1$ for $p=0.$}
 \label{fig:rhoz}
\end{figure}

To express the observables in terms of the conserved density $\rho_0,$ the above Eqs. \eqref{eq:rhom}-\eqref{eq:Vpm} are to be supplemented with
the density-fugacity relation
\bea
\rho_0(z) = \frac{z}{L} \frac{\id}{\id z} \log Z_L \simeq z \frac{\id}{\id z} \log \lambda_+. \label{eq:rhoz}
\eea
In the thermodynamic limit $L \to \infty,$ the canonical system with a fixed density  of vacancies $\rho_0$ corresponds to a particular value of the fugacity which is obtained by solving Eq. \eqref{eq:rhoz}. The invertibility of the above relation to obtain a unique fugacity $z=z(\rho_0)$ for any density $\rho_0$ guarantees the equivalence of the canonical and grand canonical ensembles. Figure \ref{fig:rhoz} shows plots of $\rho_0(z)$ versus $z$ for different values of $p$ for a fixed $\alpha,$ which is a smooth function of $z$ for any $p>0.$ 

\subsection{Phase separation transition: $p=0$}

\begin{figure}[th]
 \centering
 \includegraphics[width=8.5 cm]{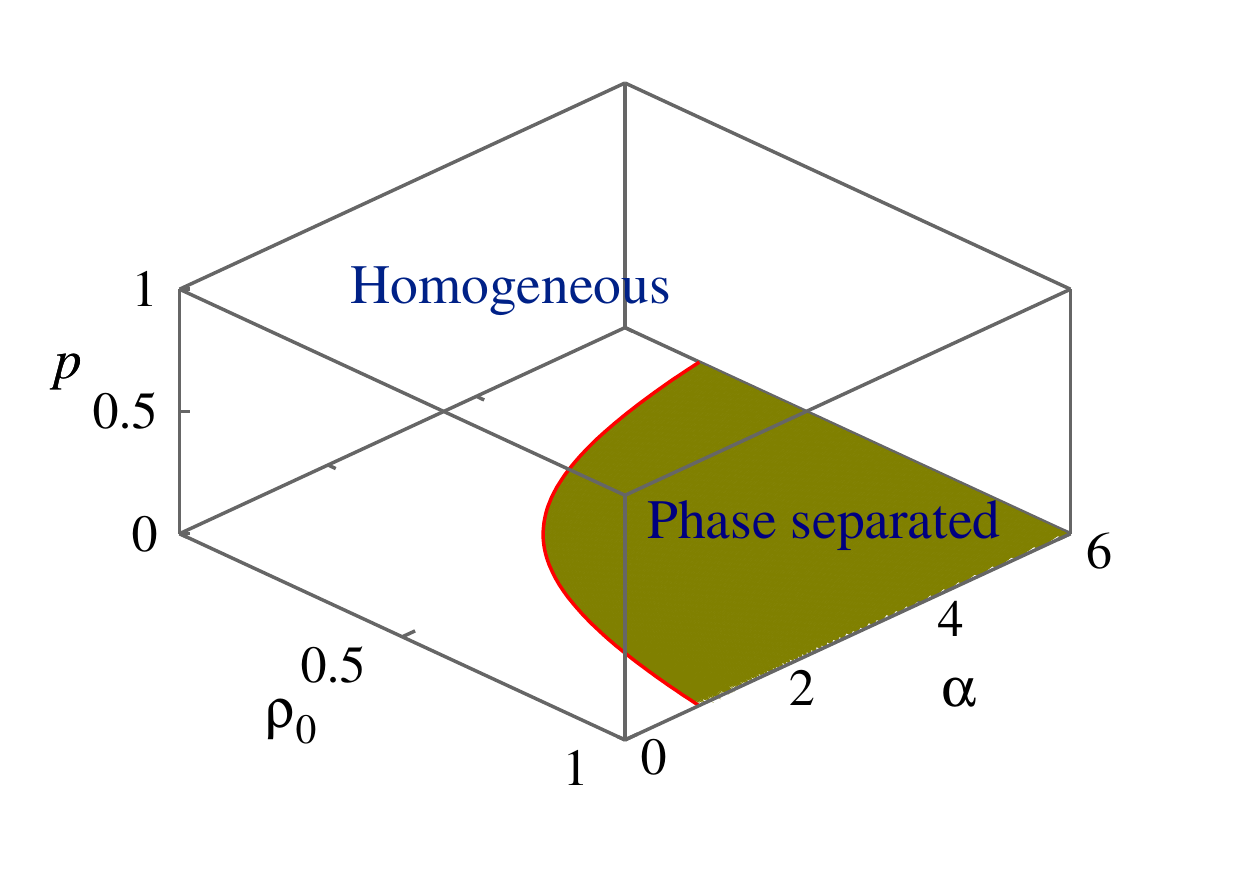}
 \caption{(Color Online) Phase diagram: The phase separation transition occurs in the $p=0$ plane only. The critical line (red curve) separates the phase separated region (green shaded region) from the homogeneous one in the $\rho_0$ -- $\alpha$ plane; there is no transition for $\alpha \le 1.$}
 \label{fig:ph}
\end{figure}

Let us focus on the $p=0$ plane. In that case the eigenvalues $\lambda_{\pm}$ take the simple form
\bea
\lambda_+ = 1+z, \quad \lambda_-= \alpha z  
\eea
and cross each other at  $z^*={1 \over \alpha -1}$ for any $\alpha > 1.$ Crossing of eigenvalues is a standard signature of the presence of a singularity in the system which is reflected as a discontinuity in the density-fugacity relation
\bea
\rho_0(z)= \left \lbrace 
\begin{split}
{z \over 1+z} & \quad {\rm for} \;\; z \le z^*, \cr
1 & \quad {\rm for} \;\; z>z^* .
\end{split}
\right.
\label{eq:rho_p0}
\eea
Clearly, for a canonical system with fixed density larger than $\rho_0(z^*) = 1/\alpha$ no grand canonical correspondence is possible. This breaking of ensemble equivalence indicates a transition --- for $p=0$ there is a phase separation transition when the density $\rho_0$ is increased beyond its critical value $\rho_c = 1/\alpha$ for any $\alpha>1$. The phase separated state exists only on the $p=0$ plane, bounded by the critical line $\rho_0 = 1/\alpha$ (see the phase diagram in Fig. \ref{fig:ph}). In fact, no transition is possible for any finite $p>0$ or $\alpha \le 1$ since the eigenvalues $\lambda_{\pm}$ of $T$  cannot cross for any $p>0$ or $\alpha \le 1$ and there the system is always in a homogeneous phase. 

In the low-density regime, i.e., for $\rho_0 < \rho_c,$  the stationary state becomes particularly simple for $p=0.$  
Let us remember that $p=0$ implies $p_+=0$ --- no negative particles are created, but they can convert to positive particles with rate $p_-$ and hence the number of negative particles can only decrease. Consequently, as is clearly seen from Eqs. \eqref{eq:rhom} and \eqref{eq:Nmav} (recall that $\lambda_+= 1+z$ here), both $\rho_-$ and $\la N_- \ra$ vanish in the stationary state for any density $\rho_0 < \rho_c.$  

The particle correlations $V_+(r)$ and $V_{+-}(r)$  (from Eqs. \eqref{eq:V+} and \eqref{eq:Vpm}), also vanish for $p=0$. In fact, it is easy to see that all spatial correlations become zero on this plane. This is not surprising since when negative particles are absent, the dynamics is identical to that of ASEP on a ring with a single species particle density $\rho_+ = 1-\rho_0,$ which is known to have no spatial correlations in the thermodynamic limit. The average current of the positive particles also takes the usual ASEP form \cite{asep2},  for $p=0$
\bea
J_+ &=& \alpha_+ \la +0 \ra 
= \frac {\alpha_+}{Z_L}Tr[DAT^{L-2}] \cr
&=&   \alpha_+ (1-\rho_0)\rho_0. \n
\eea 

It is still useful to look at the correlation length $\xi,$ as defined in Eq. \eqref{eq:xi}; for $p=0$
\bea
\xi = - \left[\log \frac{\alpha z}{1+z} \right]^{-1} = - [\log \alpha \rho_0 ]^{-1}
\eea
where Eq. \eqref{eq:rho_p0} has been used in the last step. Near $\rho_c=1/\alpha$ 
\bea
\xi \sim \left(\rho_c -\rho_0\right)^{-1}. \label{eq:xi0}
\eea
On the $p=0$ plane, the phase transition is thus associated with a diverging length scale as the critical line is approached from the low-density regime.

In the high-density i.e., $\rho_0 > \rho_c$ regime, the ensemble equivalence breaks down and the phase separated state cannot be described within this formalism. We have performed Monte Carlo simulations to investigate this regime which shows that for a large enough system there is typically a single negative particle surviving (see Fig. \ref{fig:evolution} for a typical snapshot). Even though  the macroscopic density $\rho_- \simeq 1/L$ still vanishes in the thermodynamic limit the system goes to a very different state in this regime.  A domain of vacancies of macroscopic size $\la N_- \ra$ (defined in Eq. \eqref{eq:Nmav}) forms in front of this particle;  the rest of the system is expected to remain homogeneous with vacancy density $\rho_c.$ This in turn implies,  for $\rho_0 > \rho_c,$ the average fraction $\kappa$ of sites occupied by the domain,
\bea 
\kappa = \frac 1L\la N_- \ra = \frac{(\rho_0 - \rho_c)}{(1-\rho_c)}, \label{eq:C0}
\eea
which increases linearly with the disntance from the critical line. This prediction is verified in Fig. \ref{fig:mn_rho}(b) where the domain size $\kappa $  obtained from the numerical simulations  of a system of size $L=10^4$ (blue symbols) is plotted as a function of $\rho_0$  which matches excellently with the prediction of the above equation (solid line).  

 A comment is in order about the effect of the finite system size on this phenomenon. The formation of the macroscopic domain is stable only when $\la N_- \ra \gg 1.$ In other words, to observe the macroscopic domain of vacancies for a fixed density $\rho_0,$ the system size $L$ must be large enough so that $L  \gg (1-\rho_c)/(\rho_0 - \rho_c)$ is satisfied (from Eq. \eqref{eq:C0}).

Later, in Sec. \ref{sec:cano} we will revisit the phase separated state within the canonical formulation and discuss its connection with a condensate in an equivalent zero-range process.

\subsection{Approach to $p=0$ plane: Off-critical behaviour}

The parameter $p$ can be thought of as another control parameter and it is instructive to look at the behaviour of observables as $p$ approaches the critical value $0,$ keeping the density fixed. In fact, as we will see below, the $p$ direction provides access to the non-trivial aspects of the critical behaviour.

Moving away from the $p=0$ plane the ensemble equivalence is restored, and all the static observables can be calculated analytically. The strategy is the same as before,  using the solution $z=z(\rho_0;p)$ of Eq. \eqref{eq:rhoz}, both $\rho_-$ and $\la N_-\ra,$ as defined in Eqs. \eqref{eq:rhom} and \eqref{eq:Nmav}, are obtained as a function of $\rho_0$ and $p.$

\begin{figure}[t]
 \centering
 \includegraphics[width=8.8 cm]{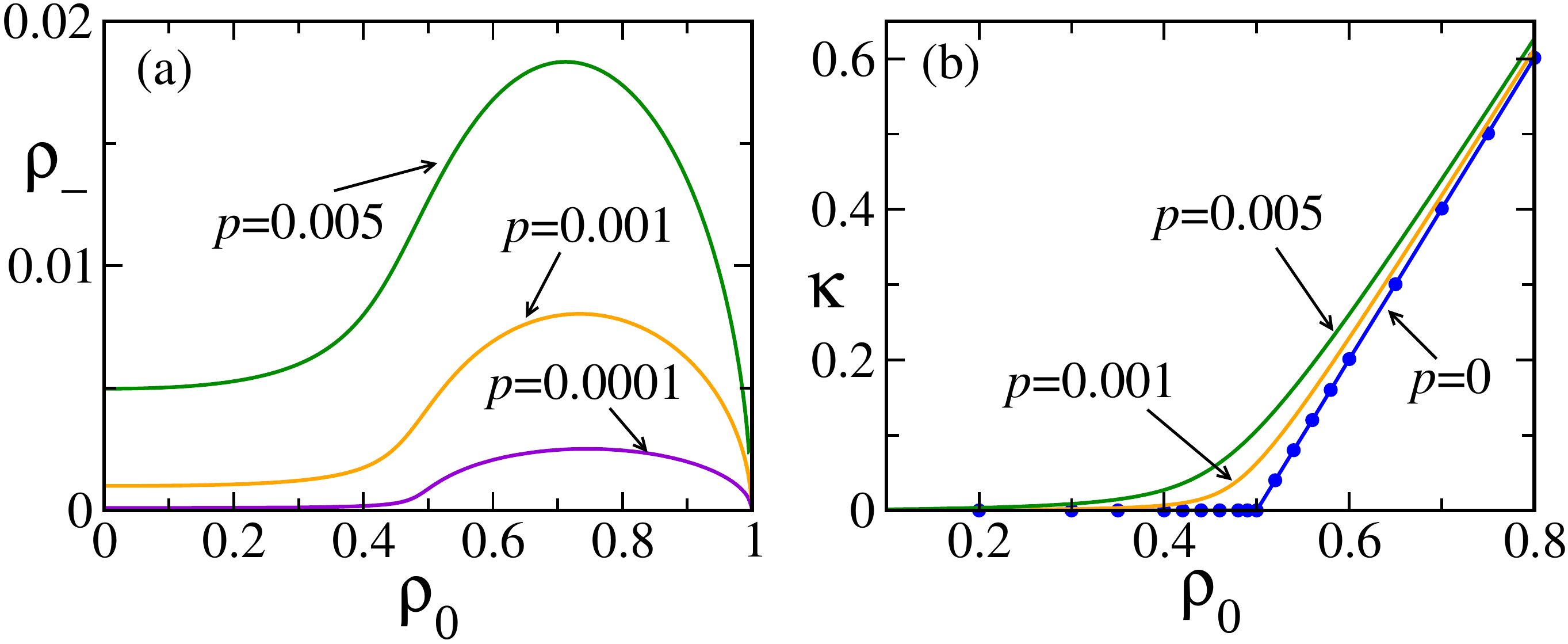}
 \caption{(Color Online) (a) The density of negative particles $\rho_-$ versus $\rho_0$ for different values of $p.$ (b) Domain size $\kappa $ versus $\rho_0$ for different values of $p.$ For $p=0$ the domain size obtained from Monte Carlo simulation (blue symbols) of a system of size $L=10^4$ is compared with the prediction in Eq. \eqref{eq:C0} (solid line). Here $\alpha=2.$ }
 \label{fig:mn_rho}
\end{figure}

Figure \ref{fig:mn_rho}(a) shows a plot of $\rho_-$ as a function of $\rho_0$ for different (small) values of $p;$ it increases slowly up to $\rho_0 \sim \rho_c$ and shows sharp growth after that.  Figure \ref{fig:mn_rho}(b) shows the same plot for $\kappa$ which remains vanishingly small up to the critical density and increases linearly thereafter.

The qualitatively different nature of the two regimes is also reflected in the approach to the $p=0$ plane. 
Below the critical density $\rho_c,$ the system is regular, both the density of negative particles $\rho_-$ and  $\kappa$ vanish linearly as $p \to 0.$

At the critical density, i.e., for $\rho_0 =\rho_c,$ both $\rho_-$ and  $\kappa$ show singular behaviour, but with different exponents. A series expansion around $p=0$ yields, to the leading order,  
\bea
\rho_-(p) &\sim & \frac{(\alpha -1)}{\alpha^{4/3}} p^{2/3},  \cr
\kappa(p) &\sim & \frac{1}{\alpha^{2/3}} p^{1/3}. \n
\eea
The average domain size of vacancies per negative particle $\kappa/\rho_- \sim p^{1/3}.$   

Yet a different behaviour is seen in the high-density regime $\rho_0 > \rho_c;$ to the leading order,
\bea
\rho_-(p) \sim  \sqrt{\frac{(\alpha \rho -1)(1- \rho)}{\alpha}} \sqrt{p}. \n
\eea

Figures \ref{fig:mn_rho}(a) and (b) show plots of $\rho_-$ and $\kappa $ at the critical point, below and above it. The series expansion for $\kappa $ could not be done for $\rho_0 > \rho_c.$ 
However, exact numerical value of $\kappa$ can be computed for any $p>0$ using Mathematica from Eqs. \eqref{eq:rhoz} and \eqref{eq:Nmav}. This is plotted for a fixed $\rho_0 > \rho_c$ (green diamonds) in Fig. \ref{fig:mn_rho}(b) and strongly suggests,
\bea
\kappa(p) - \kappa(0) \sim p^{1/2} \n
\eea
$\kappa(0)$ being given by Eq. \eqref{eq:C0}. In fact, $\kappa$ shows same behaviour for any $\rho_0$ in this high-density regime.

\begin{figure}[t]
 \centering
 \includegraphics[width=8.7 cm]{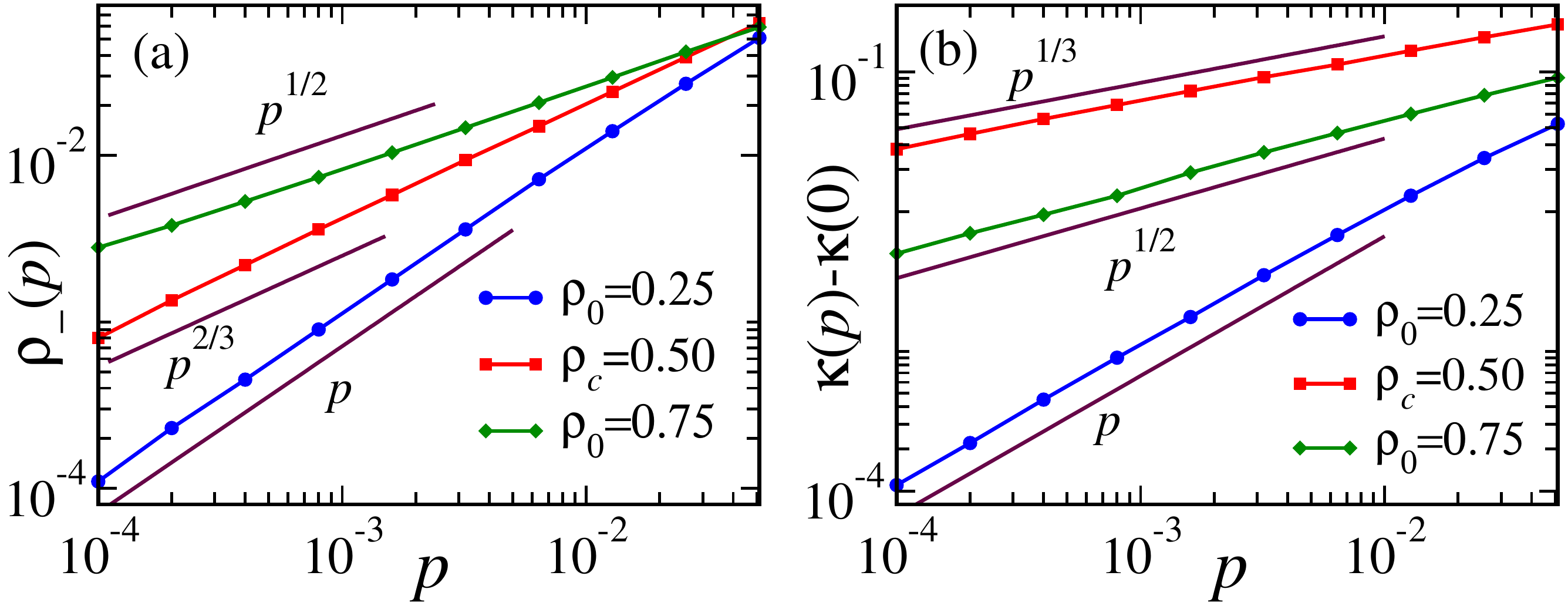}
 \caption{(Color Online) Off-critical behaviour: The density of negative particles $\rho_-$ (a) and average cluster size $\kappa$ (b) plotted against $p$ for three different values of $\rho_0$ $i.e.$ below, at and above critical density $\rho_c=0.5$   
  }
 \label{fig:mn_p}
\end{figure}

To summarize, near the plane $p=0,$ for fixed $\rho_0,$ the density of negative particles vanishes with a critical exponent $\beta,$  
\bea
\rho_-(p) \sim p^\beta , 
\eea
where  
\bea
\beta = \left \{ \begin{split}
            2/3 \quad {\rm for} \;\; \rho_0 = \rho_c, \cr
            1/2 \quad \; {\rm for} \;\; \rho > \rho_c.       
               \end{split}
 \right. \n
\eea
Surprisingly, the singular behaviour is present even deep inside the high-density phase, far away from the critical point. 
 Note that, for a finite system size $L,$ this behaviour can only be observed for $p \gg L^{-1/\beta}.$  Otherwise, negative particles do not survive, and the system becomes homogeneous.

\begin{figure}[t]
 \centering
 \includegraphics[width=8.7 cm]{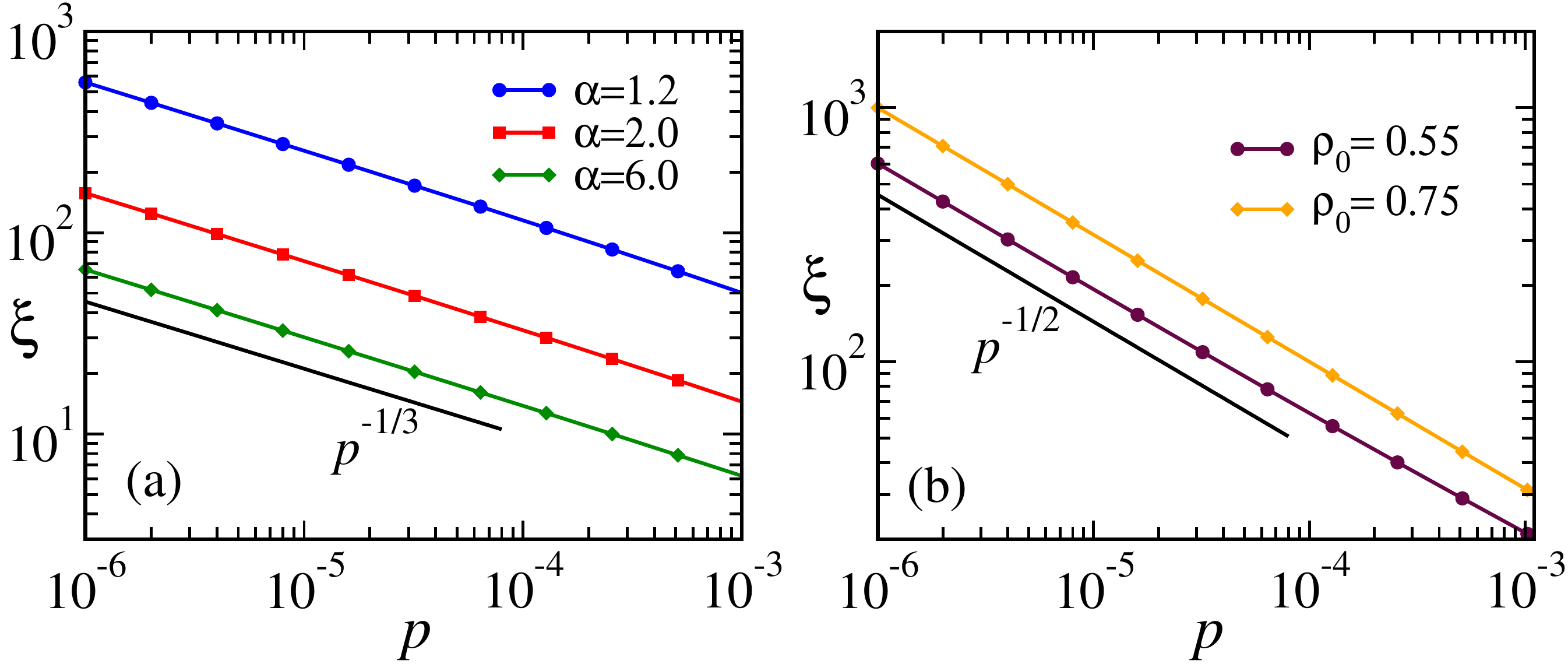}
 \caption{(Color Online) Correlation length $\xi$ plotted as a function of $p:$ (a) At the critical point $\rho_c = 1/\alpha$ for three different values of $\alpha.$ $\xi$ diverges with exponent $\nu_\perp = 1/2.$
(b) For two different densities $\rho_0$ above the critical density for $\alpha=2.$ In this case $\nu_\perp =1/2.$
 The solid lines provide a guide to the eye for the corresponding power laws.  }
 \label{fig:xi}
\end{figure}

To gain more insight about the phase separated regime we take a look at the correlation length, as defined in Eq. \eqref{eq:xi}. 
At $\rho_0 = \rho_c$, for small $p,$ 
\bea
\frac{\lambda_-}{\lambda_+} = 1 - \frac{\alpha -1}{\alpha^{2/3}} p^{1/3} + {\cal O}(p^{2/3}).
\eea
This in turn implies, as $p \to 0,$ the spatial correlation length  diverges as,
\bea
\xi \sim p^{-1/3}. \n
\eea
The associated critical exponent is thus $\nu_\perp  = 1/3.$ The exact numerical values of $\xi,$ as obtained from Eq. \eqref{eq:xi} are plotted in Fig. \ref{fig:xi}(a) as a function of $p,$ for different values of $\rho_c=1/\alpha.$ 

The analytical expansion is not possible for the high-density regime. However, as before, for any $p>0$ the numerical value of $\xi$ can be calculated using  Eqs. \eqref{eq:xi} and \eqref{eq:rhoz}. This is plotted in Fig. \ref{fig:xi}(b) for different values of $\rho_0$ above the critical value and suggests that
\bea
\xi \sim p^{-1/2}, \n
\eea
i.e., $\nu_\perp=1/2$ for $\rho_0 > \rho_c,$ different from that at the critical point.

 It appears that the whole phase separated regime is `critical': for all densities above $\rho_c$ the system is associated with a diverging correlation length as $p \to 0$. Moreover, the set of critical exponents characterizing the phase separated regime are different from those on the boundary line $\rho_0 = 1/\alpha.$

\subsection{Canonical Ensemble: Mapping to a Zero-Range Process}\label{sec:cano}

The study of the actual phase separated state is not possible within the grand canonical scheme since the ensemble equivalence breaks down. However, it is also possible to study the phenomenon directly within the canonical ensemble using a mapping of the exclusion dynamics  to a zero-range process which we discuss in this section.

A zero-range process (ZRP) describes stochastic motion of particles on a lattice where each site, also referred to as a box, can accommodate an arbitrary number of particles and the hopping rate depends on the number of particles in the departure box only \cite{zrp}. The stationary state for any generic ZRP has a factorized form where the lattice sites become independent. 

The exclusion dynamics \eqref{eq:model} discussed in  Sec. \ref{sec:model} can be mapped to a ZRP with constant particle hopping rates by identifying positive and negative particles as two different kinds of boxes $\tau=\pm$ respectively \cite{twobox}. 
An uninterrupted sequence of $n$ number of $0$s on the lattice to the right  of  a $+(-)$ particle denotes a   $+(-)$ box containing $n$ particles (see Fig. \ref{fig:cartoon} for a schematic representation). In this ZRP picture particles from a $+$ (respectively $-$) box hop to the left box with a constant rate  $\alpha_+$ (respectively $\alpha_-$) and an empty $+$ (or $-$) box can alter its state with rate  $p_+$ (or $p_-$).
Note that even though the total number of boxes $M$ and particles $N_0$ is conserved in the ZRP, the  number of boxes of each kind and particles in them can fluctuate.

Following the result of Ref.~\cite{twobox}, the stationary state weight of a generic configuration $\{ n_i \tau_i \}$ in this ZRP has a product structure,
\bea
P(\{ n_i \tau_i \}) \sim \prod_{i} f_{\tau_i}(n_i) \n
\eea
with $f_{\tau_i}(n)$ being the weight that the $i^{th}$ box in the state $\tau_i$ contains $n$ particles. It turns out that,
\beq
f_+(n) = 1; ~~ f_-(n) = p \alpha^n  \n
\eeq
where, of course, only the ratios of the flip rates $p$ and hopping rates $\alpha$ appear. 


 Writing the formal canonical partition function is now straightforward, 
\bea
Z_{L,N_0}  &=& {L \over L-N_0} \sum_{N_-=0}^{N_0} \sum_{M_-=0}^{L-N_0} 
 C_{N_+}^{N_+ +M_+-1} \cr
&& \qquad \qquad C_{N_-}^{N_-+M_--1} C^{L-N_0}_{M_+} p^{M_-} \alpha^{N_-} \cr
&& \n
\eea
where $N_{+}$ ($N_{-}$) denotes the total number of particles in positive (negative) boxes with $N_+ + N_-=N_0.$
The combinatorial factors count the number of ways this partitioning can be done. 
Since the transition occurs for $p = 0$, and the number of negative box becomes ${\cal O}(1)$ in that limit, an even simpler scenario can be used to capture the basic physical picture. We take the case where there is only one negative particle in the system, i.e.,  $M_-=1$ which, as we will see below, turns out to be  enough to explore the condensate phase. 

The canonical partition function for this case reduces to,
\bea
Z_{L,N_0} = (L-N_0)\sum_{N_-=0}^{N_0} C^{L-N_--2}_{N_0-N_-}\alpha^{N_-}. \n
\eea
Note that this partition function corresponds to a system where the second part of the dynamics \eqref{eq:model} is absent, i.e., boxes cannot change their type. So, the above partition function corresponds to a specific conserved sector of the original configuration space with a single negative particle and $M-1$ positive particles.

This is nothing but a disordered ZRP with a single defect and it is already known that a condensation transition can occur if the defect box has a smaller hopping rate than the other boxes\cite{zrp}. This condition translates to $\alpha > 1$ in the present case, same as what we have obtained in Sec. \ref{sec:exact}.    
 
The asymptotic behaviour of $Z_{L,N_0}$ in the limit $L,N_0 \to \infty$ but fixed $\rho_0=N_0/L$ can be investigated
using the method of steepest descent \cite{zrp},
\bea
Z_{L,N_0} &\simeq & \left\{
\begin{split}
 (L-N_0) C^L_{N_0} {(1-\rho_0)^2\over (1-\alpha \rho_0)} \;\; {\rm for}\;  \rho_0<{1 \over \alpha}, \;\; \cr
		 (L-N_0) {\alpha^{L-1} \over (\alpha -1)^{L-N_0-1}} \;\; {\rm for}\; \rho_0>{1 \over \alpha}. 
\end{split}
\right. \n
\eea 
Clearly, this conserved system undergoes a phase transition at $\rho_0=1/\alpha,$ same as the original system at $p=0$.
For $\rho_0 > 1/\alpha,$ a macroscopic condensate is formed in the negative box --- equivalent to the domain forming in front of the negative particle in the exclusion process.

The size of the condensate, which corresponds to the domain size in the exclusion picture, is nothing but the average number of particles contained in the negative box, and turns out to be
\bea
\la N_- \ra &=& \alpha \frac{\id}{\id \alpha} \log ~Z_{L,N_0} \cr
 &=& \left\{
 \begin{split}
  \frac{\alpha \rho_0}{1- \alpha \rho_0} \quad{\rm for}\quad \rho_0 < \rho_c, \cr
 L \frac{(\alpha \rho_0 -1)}{(\alpha-1)} \quad {\rm for}\quad \rho_0 > \rho_c.
 \end{split}
 \right. \n
\eea
This estimate, as expected, is identical to the previous result Eq. \eqref{eq:C0} in the condensate phase. Note that, in contrast to the original dynamics, the negative box is still present even below the critical density and hence $\la N_-\ra$ gets a non-zero value in this `reduced' ZRP.

 \begin{table}
\begin{tabular}{c||ccccc}
\hline
~~ & ~~~ $\beta$ ~~~& ~~~$\nu_\perp$ ~~~ & ~~~$\delta$~~~ & ~~~$z$~~~ \cr
\hline\hline
$\rho_0=\rho_c$ & $2/3$ & $1/3$  & $1$ & $1$ \cr
$\rho_0 > \rho_c$ & $1/2$ & $1/2$ & $1/2$ & $2$ \cr
\hline\hline
\end{tabular}
  \caption{Summary of the critical exponents associated with the phase separation transition.}
  \label{tab:exp} 
 \end{table}

On the other hand, the prediction from the single defect ZRP is expected to work for any value of density for observables which do not involve the negative particle in the original process. For example, the current of positive particles is given by 
\bea
J_+ &=& \alpha_+ \la +0 \ra \cr
&=& \alpha_+{(1-\rho_0) \over Z_{L,N_0}} (L-N_0-1) S, \label{eq:Jp} \\
{\rm where}\quad \qquad S &=& \sum_{N_+ =1}^{N_0}~ C^{L-N_0-3 + N_+}_{N_+ -1} \alpha^{N_0-N_+}. \n
\eea
As before, the combinatorial factors correspond to the number of ways the particles can be distributed under the relevant conditions. 

\begin{figure}[th]
 \centering
 \includegraphics[width=5.8 cm]{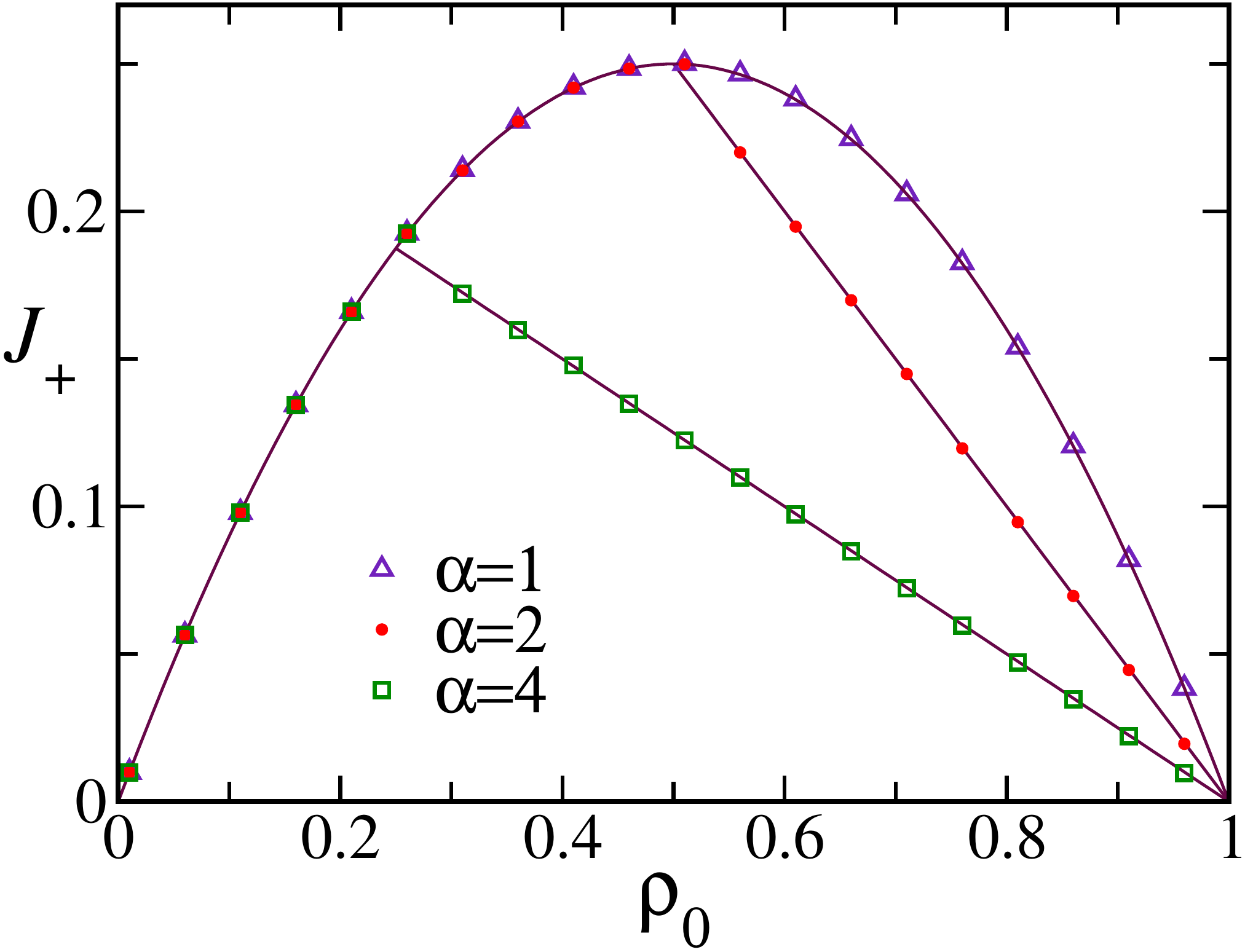}
 \caption{(Color Online) Current of positive particles $J_+$ for $p=0,$ as a function of density $\rho_0$ for different values of $\alpha=1,2$ and $4,$ obtained from numerical simulation of a system of size $L=10^4.$ The hopping rate for positive particles $\alpha_+ =1$ here. }
 \label{fig:cur}
\end{figure}

Once again, we use the method of steepest descent to compute the asymptotic behaviour,
\bea
S \simeq \left \{ 
\begin{split}
  C^L_{N_0} \frac{\rho_0}{1- \alpha \rho_0}(1-\rho_0)^2 \quad{\rm for}\; \rho_0 < \frac 1\alpha, \cr
 \frac{\alpha^{L-2}}{(\alpha-1)^{L-N_0-1}} \quad \quad \quad {\rm for} \; \rho_0 > \frac 1\alpha. 
\end{split}
\right. \n
\eea

Combining the above equation with Eq. \eqref{eq:Jp}, we get, 
\bea
J_+ =\left \{ 
\begin{split}
 \alpha_+ \rho_0(1- \rho_0) \quad {\rm for} \; \rho_0 < {1 \over \alpha}, \cr
\frac{\alpha_+}{\alpha} (1- \rho_0) \quad \;\, {\rm for} \; \; \rho_0 \ge {1 \over \alpha}. 
\end{split} \label{eq:Jzrp}
\right.
\eea
The form of the current is same as in ASEP for the homogeneous phase; in the condensation regime it decreases linearly with density. Note that the current is a continuous function of the density $\rho_0$ but its derivative is discontinuous at $\rho_0=1/\alpha.$

Figure \ref{fig:cur} shows plot of $J_+,$ as obtained from numerical simulation for $p=0,$ for different values of $\alpha.$ For $\alpha=1,$  there is no transition and the current follows the ASEP form for all values of density. In fact, this is true for any $\alpha \le 1.$ For $\alpha > 1$ the current exactly matches the prediction in Eq. \eqref{eq:Jzrp}, with a sharp change in behaviour across $\rho_c$. This provides an additional confirmation that the reduced ZRP picture with a single defect box adequately describes the phase separated state that appears in the $p=0$ limit.

\section{Relaxation Dynamics}\label{sec:dyn}

For a complete characterization of the phase transition one also needs to explore the dynamical behaviour of the system for which we take recourse to Monte Carlo simulations. The temporal decay of  $\rho_-(t)$ is measured starting from a homogeneous configuration with equal numbers of positive and negative particles for different values of $p > 0$ at and above the critical density $\rho_c.$ 

\begin{figure}[t]
 \centering
 \includegraphics[width=8.6 cm]{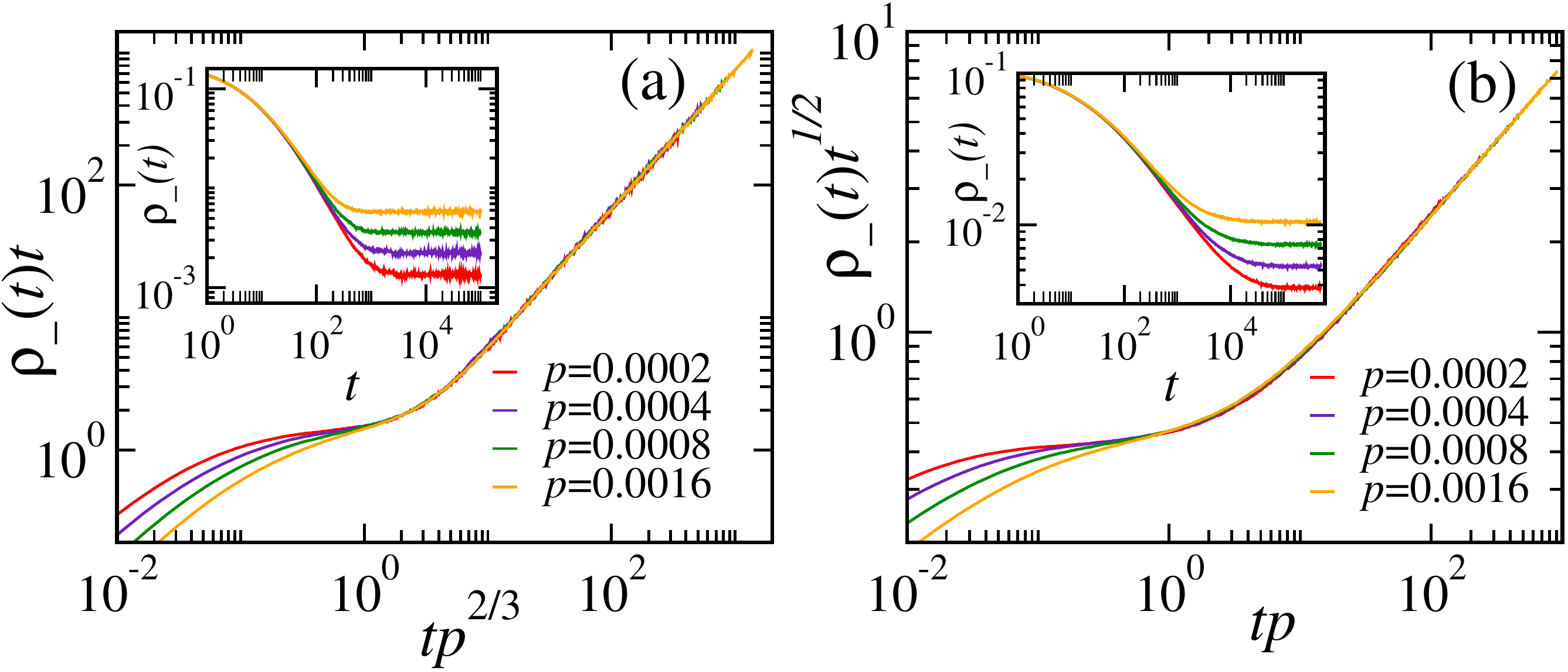}
 \caption{(Color Online) Scaling collapse for the density of negative particles $\rho_-(t)$ as a function of $tp^{\nu_\|}$ for different values of $p.$ (a) At $\rho_0 = \rho_c:$ corresponding $\nu_\|=2/3, \delta =1.$ (b) $\rho_0 > \rho_c:$ corresponding $\nu_\|=1/2, \delta =1/2.$ The insets show the unscaled data in both cases; value of $p$ increases from upper to lower curves. Here $\alpha=2$ so that $\rho_c =0.5.$ The system size $L=10^3.$ } \label{fig:scaling}
\end{figure}

At the critical point ($p=0, \rho_0=1/\alpha$) the density $\rho_-(t)$ is expected to show a power law decay with some exponent $\delta$. For small $p>0$ at the critical density $\rho_-$ saturates to the stationary value $\sim p^\beta$ and the phenomenological scaling form can be expressed as,
\bea
\rho_-(t) = t^{-\delta} {\cal F}(tp^{\nu_\|}), \label{eq:scaling}
\eea
where $\nu_\| =\beta / \delta$ is another critical exponent characterizing the temporal correlation length
and ${\cal F}$ is a scaling function. In fact, for this system, a similar scaling form is expected to hold even deep inside the high-density regime since both $\rho_-$ and $\xi$ show signatures of criticality for any density above $\rho_c,$ possibly with a different exponent and a different scaling function.

Figures \ref{fig:scaling}(a) and (b) show the scaling collapse of the numerical data according to Eq. \eqref{eq:scaling} for $\rho_0 =\rho_c$ and $\rho_0 >\rho_c$ respectively. The best data collapse occurs for $\delta = 1, \nu_\|=2/3$ at the critical point $\rho_c$ and $\delta=1/2, \nu_\|=1$ for $\rho_0>\rho_c$ ---  the dynamical behaviour is distinctly different in the high-density regime. 

\subsection*{Finite size scaling}  
Finally, to investigate the finite size scaling behaviour of the system we measure the temporal decay of $\rho_-(t)$  for different values of the system size $L$ at $p=0.$ The phenomenological scaling form, in this case, is given by
\bea
\rho_-(t) = t^{-\delta} {\cal G}(t/L^z) \n
\eea
where $z$ is the dynamical exponent ${\cal G}$ is the finite size scaling function. 
The same homogeneous initial condition is used here.

\begin{figure}[t]
 \centering
 \includegraphics[width=8.8 cm]{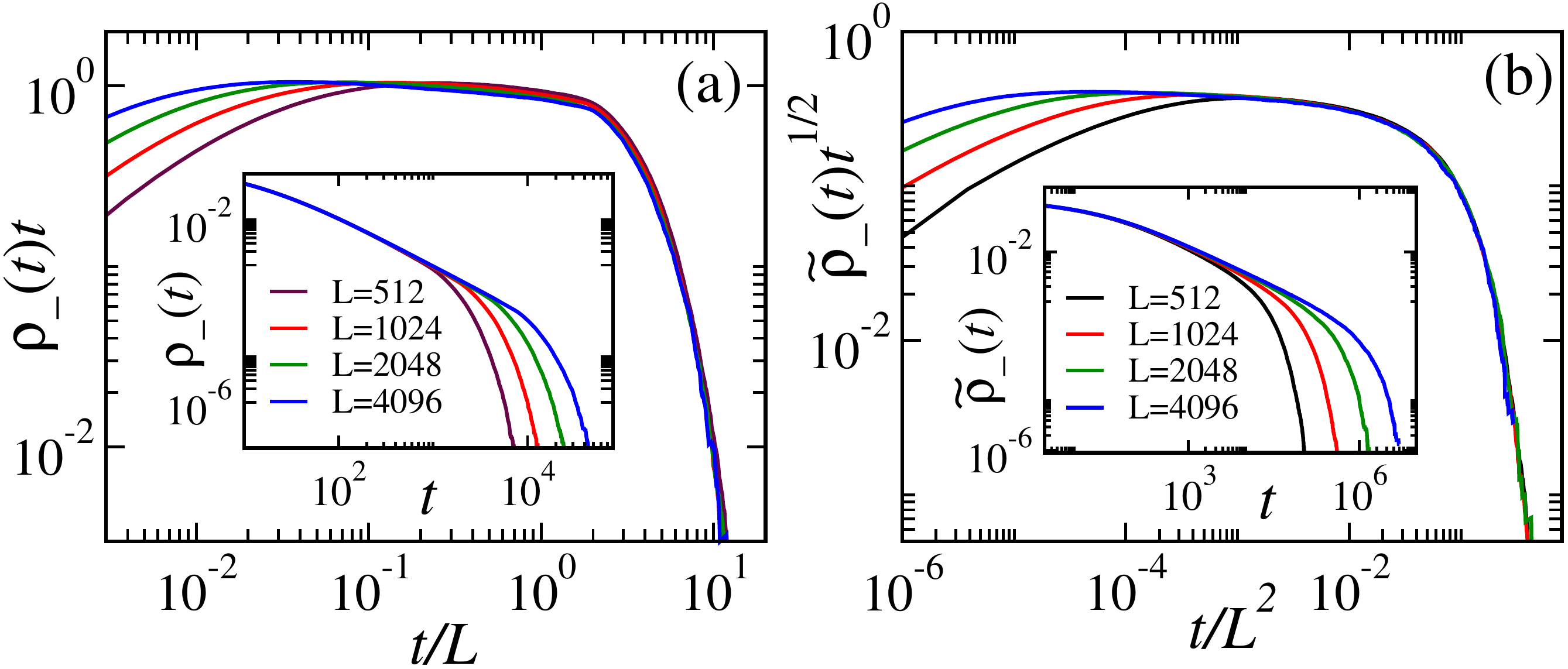}
 \caption{(Color Online) Finite size scaling at $p=0$: (a) Plot of $\rho_-(t)t^{\delta}$ versus $t/L^z$ at the critical point $\rho_c=0.5$ for different lattice sizes $L=2^9$ to $2^{12},$ the lowest curves corresponding to the smallest system size.  
The best data collapse is obtained for $z=1.$ (b) Similar plot in the high-density regime for $\rho_0=0.75$ using $\tilde \rho_-=\rho_- - 1/L;$ in this case $z=2.$ Insets show the unscaled data. Here $\alpha=2.$ }\label{fig:Lscaling}
\end{figure}

Figures \ref{fig:Lscaling}(a) and (b) show the scaling collapse of the numerical data according to the above scaling form for $\rho_0 = \rho_c$ and $\rho_0 > \rho_c$  respectively. For the latter case we have used $\tilde \rho_-(t) = \rho_-(t) -1/L$ since one negative particle survives in the stationary state.  In this regime the best collapse is obtained for $z=2.$ At the critical density $\rho_c=1/\alpha,$ however, the scaling collapse clearly suggests $z=1$  (see Table \ref{tab:exp} for a summary of the exponents).

\section{Conclusion}\label{sec:concl}

We study a phase separation transition in a one dimensional exactly solvable driven diffusive system with two species of particles, referred to as positive and negative. The dynamics does not conserve number of particles of individual species but the total number of particles remains fixed. It is shown that, in the limiting case $p=0,$ when negative particles are not allowed to be created, the system shows a transition from a homogeneous state to a phase separated one when the total density is changed. The phase separated state is characterized by presence of a microscopic number of negative particles and formation of macroscopic domains of vacancies in front of these.

Surprisingly, the phase separated state appears to be always critical, associated with a diverging correlation length for any density above the critical value $\rho_c.$ The corresponding exponents are obtained by studying the behaviour of the system in the limit $p \to 0$  for a fixed density $\rho_0$.  Both the static and the dynamic exponents are obtained from the exact solution and Monte-Carlo simulations. These exponents turn out to be different when $\rho_0=\rho_c$ and $\rho_0 > \rho_c.$

 This scenario is similar to a continuous phase transition  where the terminal point of a critical line separating two phases shows a universal behaviour which is different than on the line. Some  examples in  the non-equilibrium  context are absorbing state transitions observed in Domany-Kinzel model \cite{DK} and self-organized criticality in the sticky grain model \cite{sticky}  where the critical line showing the directed percolation behaviour ends at a special fixed  point, namely compact directed percolation. Usually, a  different critical behaviour at  the end-point is  the outcome of  an additional symmetry (particle-hole symmetry in the above  examples). In the  present study, however,  it is not clear what  is the underlying feature that makes the critical behaviour different at $\rho=\rho_c$ from that in the high-density regime.

The model can be mapped to a zero-range process and the phase separation is nothing but a condensation transition in this picture. The connection between exclusion process and zero-range process has been exploited in deriving a general criterion for phase separation transition for driven diffusive systems \cite{Kafri2002}. This usual classification of phase separation using mapping to single species ZRP relies on the generic condition of condensation transition in ZRP which does not occur for constant hopping rates. This is not the case here, however  -- the corresponding zero-range process is a disordered one where phase transition can occur even with constant rates \cite{zrp, Jain}.

The examples of phase separation transition in one dimensional nonequilibrium systems known so far are either fully conserved, and particles (irrespective of species) form large domains or with a single second class particle.  Or, in the case of non-conserved dynamics, particles of both species form separate domains. 
The model studied here provides a different example of an exactly solvable driven diffusive system where particle number for each species can fluctuate and the vacancies form the macroscopic domain.

\begin{acknowledgements}
 The author thanks P. K. Mohanty and Andrea Gambassi for many useful discussions and careful reading of the manuscript. The financial support by the ERC under Starting Grant 279391 EDEQS is acknowledged.
\end{acknowledgements}


\begin{thebibliography}{99}
\bibitem{zia} B. Schmittmann and R. K. P. Zia, {\it Statistical Mechan-
ics of Driven Diffusive Systems}, Phase Transitions and
Critical Phenomena Vol. 17, edited by C. Domb and J. L. Lebowitz (Academic, London, 1995).

\bibitem{marro} J. Marro, R. Dickman, {\it Nonequilibrium Phase Transitions in Lattice Models}, (Cambridge University Press, New York, 1999).  

\bibitem{asep1} T. M. Liggett, {\it Stochastic Interacting Systems: Voter, Contact
and Exclusion Processes}, (Springer, New York, 1999).  

\bibitem{asep2}G. M. Sch\"{u}tz, in Phase Transitions and Critical Phenomena Vol. 19, edited
by C. Domb and J. L. Lebowitz (͑Academic Press, London, 2000). 

\bibitem{asepexact} B. Derrida, E. Domany, D. Mukamel, J. Stat. Phys. {\bf 69}, 667(1992). 

\bibitem{derrida} B. Derrida, S.A. Janowsky, J.L. Lebowitz, E.R. Speer, 
J. Stat. Phys. {\bf 73}, 813 (1993).

\bibitem{evans} M.R. Evans, Europhys. Lett. \textbf{36}, 13 (1996). 

\bibitem{Krug} J. Krug and P. A. Ferrari,  J. Phys. A: Math. Gen. {\bf 29} L465 (1996).

\bibitem{ABC}  M. R. Evans, Y. Kafri, H. M. Koduvely, and D. Mukamel, Phys. Rev. Lett., {\bf 80}, 425 (1998).

\bibitem{Krug2} J. Krug, Braz. J. Phys.,{\bf 30}, 97 (2000).  

\bibitem{rasep} Urna Basu and P. K. Mohanty, Phys. Rev. E {\bf 79}, 041143 (2009).

\bibitem {Kafri2003}  Y. Kafri, E. Levine, D. Mukamel, G. M. Sch\"{u}tz, and R. D. Willmann, \pre {\bf 68}, 035101 (R) (2003).

\bibitem{PK} A. Kundu and P. K. Mohanty, Physica A, {\bf 390}, 1585 (2011).


\bibitem{Jafarpour} F. H. Jafarpour, J. Phys. A: Math. Gen. {\bf 33}, 8673(2000).

\bibitem{Jafarpour2} M. Ghadermazi and  F.H. Jafarpour, J. Theor. Appl. Phys. {\bf 10}, 195 (2016).

\bibitem{HH} S. Zeraati, F. H. Jafarpour, H. Hinrichsen, Phys. Rev. E {\bf 87}, 062120 (2013).


\bibitem{EvansMukamel} M. R. Evans, Y. Kafri, E. Levine and D. Mukamel, 
 J. Phys. A: Math. Gen. {\bf 35}, L433 (2002).


\bibitem{twobox} U. Basu and P. K. Mohanty, Phys. Rev. E {\bf 82}, 041117 (2010). 

\bibitem{ozrp} U. Basu and P. K. Mohanty, J. Stat. Mech.: Theory Exp. L03006 (2010).

\bibitem{MPA} R. A. Blythe and M. R. Evans, J. Phys. A: Math. Theor.
{\bf 40}, R333 (2007).

\bibitem{zrp} M. R. Evans and T. Hanney, J. Phys. A: Math. Gen. {\bf 38}, R195 (2005). 



\bibitem{DK}  E. Domany and W. Kinzel, Phys. Rev. Lett. {\bf 53}, 311 (1984).


\bibitem{sticky}  P. K. Mohanty, D. Dhar, Phys. Rev. Lett. {\bf 89}, 104303 (2002).


\bibitem {Kafri2002} Y. Kafri, E. Levine, D. Mukamel, G. M. Sch\"{u}tz, and J. T\"{o}r\"{o}k, Phys. Rev. Lett. {\bf 89}, 035702 (2002). 

\bibitem{Jain}  K. Jain and M. Barma, Phys. Rev. Lett. {\bf 91}, 135701 (2003).

\end{thebibliography}
\end{document}